\documentclass[traditabstract]{./aa}
\usepackage{graphicx}
\usepackage{txfonts}
\usepackage{natbib}
\bibpunct{(}{)}{;}{a}{}{,} 

\def\ecs{erg~cm$^{-2}$s$^{-1}$}
\def\lum{erg~s$^{-1}$}
\def\bron{A~1246-58}
\renewcommand{\tabcolsep}{1.5mm}  

\begin{document}

\title{An X-ray and optical study of \\ the ultracompact X-ray binary
  \bron}

\titlerunning{X-ray and optical study of \bron}
\authorrunning{J.J.M. in 't Zand et al.}

\author{J.J.M.~in~'t~Zand\inst{1,2}, C.G. Bassa\inst{3},
P.G. Jonker\inst{1,2,4}, L.~Keek\inst{1,2}, F. Verbunt\inst{2},
M. M\'{e}ndez\inst{5} \& C.B.~Markwardt\inst{6,7}}

\offprints{J.J.M. in 't Zand, email {\tt jeanz@sron.nl}}

\institute{     SRON Netherlands Institute for Space Research, Sorbonnelaan 2,
                3584 CA Utrecht, the Netherlands 
	 \and
                Astronomical Institute, Utrecht University, P.O. Box 80000,
                3508 TA Utrecht, the Netherlands
         \and
                Physics Dept., McGill University, 3600 Rue University,
                Montreal, QC, H3A 2T8, Canada
	 \and
                Harvard-Smithsonian Center for Astrophysics, 60 Garden Street,
                Cambridge, MA 02138, U.S.A.
         \and
                Kapteyn Astronomical Institute, University
                of Groningen, P.O. Box 800, 9700 AV Groningen, the Netherlands
         \and
                Department of Astronomy, University of Maryland,
                College Park, MD 20742, U.S.A.
         \and
                Astroparticle Physics Laboratory, Mail Code 661, NASA
                Goddard Space Flight Center, Greenbelt, MD 20771, U.S.A.
          }

\date{Accepted for publication in A\&A on March 21, 2008}

\abstract{Results are discussed of an X-ray and optical observation
  campaign of the low-mass X-ray binary \bron\ performed with
  instruments on Satellite per Astronomia X ('BeppoSAX'), the Rossi
  X-ray Timing Explorer (RXTE), the X-ray Multi-mirror Mission
  ('XMM-Newton'), the {\em Swift} mission, and the Very Large
  Telescope. Spectra and flux time histories are studied. The most
  important results are the lack of hydrogen spectral features in the
  optical spectrum, supporting the proposition that this is an
  ultracompact X-ray binary (UCXB), the determination of a 4.3 kpc
  distance from time-resolved spectroscopy of thermonuclear X-ray
  bursts, and the detection of intermediately long thermonuclear
  bursts as seen in a number of other UCXBs. There is evidence for a
  Ne/O abundance ratio in the line of sight that is higher than solar
  and variable. This may be due to different changes in the ionization
  degrees of Ne and O, which may be related to the variable irradiating
  flux. We discuss the spectral variability and the peculiarities of
  the long-term light curve.

}

\maketitle 

\section{Introduction}
\label{intro}

Ultracompact X-ray binaries (UCXBs) consist of a dwarf star that is
overflowing its Roche lobe and transferring matter via an accretion
disk on an accompanying neutron star (or black hole, although such a
case has not been identified yet) in an orbit with a period shorter
than 80~min. The period threshold implies that the donor star is
hydrogen deficient \citep[e.g.,][]{nrj86} because a hydrogen-rich
donor would not fit in the Roche lobe. The discovery rate of
(candidate) UCXBs has increased in recent years. Currently, there are
9 UCXBs with measured orbital periods, 4 with tentative period
measurements, and 15 systems with other indications that they are
ultracompact \citep[e.g.,][]{nele06,zan07}. The fraction of
ultracompact systems in the group of Galactic low-mass X-ray binaries
(LMXBs) is large: presumably of the order of 50\% or more
\citep{zan07}.

The source \bron\ was discovered three decades ago
\citep{carpenter77}. Dedicated X-ray observations were sparse during
the first two decades with only one (unpublished) EXOSAT
observation. A serendipitous detection of an X-ray burst in 1997
\citep{pir97,bol97} identified \bron\ as a LMXB, with a
neutron star as accretor. In 2005 we initiated a vigorous observation
campaign of this object to obtain a more complete picture of the
properties of this LMXB, encompassing X-ray as well as optical
observations. Preliminary results have been published in
\citet{bjzv06}, who identified a $V=19.45$ optical counterpart, and in
\citet{jonker07} who found a quasi-periodic oscillation of exceptional
high frequency (1258 Hz) and the highest amplitude among LMXBs. Based
on an optical to X-ray flux ratio that is relatively low for an LMXB,
\bron\ was identified as a candidate UCXB \citep{bjzv06}.

We discuss in this paper results from the observation campaign on
\bron. The data were taken with state-of-the-art instrumentation such
as the European XMM-Newton X-ray observatory and the Very Large
Telescope.  The new data focus on X-ray bursts and on measurements of
the X-ray and optical spectra.  In Sect.~\ref{obs}, an overview is
provided of all observations. Section~\ref{results} discusses the
analysis results on the acquired data sets. Finally, in
Sect.~\ref{discussion}, the results are discussed in the context of
UCXBs and accretion disk theory.

\section{Observations and data reduction}
\label{obs}

\begin{table*}
\centering
\caption[]{Overview of all data sets on \bron.\label{tabobs}}
\begin{tabular}{lllllc}
\hline\hline
Facility   & Instrument & Wavelength/& Data modes & Dates & Exposure time\\
           &            & photon energy &         &       & \multicolumn{1}{c}{(ks)}\\
\hline
BeppoSAX   & WFC        & 2-30 keV   & Normal     & Jul 96 - Apr 02 & 8$\times10^3$\\
RXTE       & ASM        & 2-12 keV   & Normal     & Jan 96 - Jul 07 & 4.5$\times10^3$\\
           & PCA        & 2-60 keV   & Normal     & Dec 05 - Jun 07 & 104 \\
{\em Swift}& XRT        & 0.3-10 keV & WT         & Aug 11,   06    & 12 \\
XMM-Newton & EPIC-PN    & 0.6-12 keV & Fast Timing& Aug 31,   06    & 41 \\
           & EPIC-MOS   & 0.3-10 keV & Timing+Ima &                 & \\
           & RGS        & 0.6-1.6 keV& Normal     &                 & \\
           & OM         & U-band     & Fast+ima.  &                 & 36 \\
VLT        & Antu+FORS2 & 380-830 nm & Spectrosc. &Apr+May 06, Jan 07& 10 \\
\hline\hline
\end{tabular}
\end{table*}

The observation campaign comprises of X-ray and optical data from 9
instruments on 4 space-borne and 1 ground-based
facility. Table~\ref{tabobs} summarizes the observations.

\subsection{BeppoSAX}

The Satellite per Astronomia X ('BeppoSAX') observatory
\citep{boe97a}, operational between June 1996 and April 2002, carried
the Wide Field Camera (WFC) instrument package consisting of two
identical X-ray cameras each with a field of view of
40\degr$\times$40\degr\ pointing in opposite directions, a bandpass of
2 to 30 keV and a spectral resolution of 20\% full width at half
maximum (FWHM) \citep{jag97}. Except for a dedicated observation
program on the Galactic center \citep{zan04a}, the cameras were
pointed randomly at the sky. For a total of 8 Msec, \bron\ was
serendipitously in the field of view of either camera. The sensitivity
of the WFCs and the flux of \bron\ was such that the source was always
near the detection threshold and we refrain from an analysis of the
data on the persistent emission and focus on four detected X-ray
bursts.

\subsection{RXTE}

The Rossi X-ray Timing Explorer (RXTE) carries three instruments: the
All-Sky Monitor \citep[ASM; ][]{lev96}, the Proportional Counter Array
\citep[PCA;][]{jah06}, and the High-Energy X-ray Timing Experiment
\citep[HEXTE; ][]{roth98}. The ASM observes approximately 80\% of the sky each
1.5~hr orbit during one or several 90-s dwells by its three
cameras. \bron\ is far from the ecliptic, which implies that it is
covered throughout the year. The bandpass is the traditional X-ray
band between 2 and 12 keV. The source is in a relatively clean part of
the sky, which precludes source confusion and ensures optimum
sensitivity. \bron\ is fairly weak for detection with the
ASM. Therefore, the data of approximately one week of observations
need to be combined to obtain a significant signal (see
Fig.~\ref{rxte} upper panel).

Between Dec 16, 2005, and June 25, 2007, we carried out a monitoring
campaign with the PCA and HEXTE (under observation id 90042), by
performing 1~ks snapshot observations once a week. PCA and HEXTE data
are the only data that cover photon energies above 10 keV. The PCA
consists of 5 proportional counter units (PCUs) with co-aligned
collimated fields of view of 1\degr$\times$1\degr\ (FWHM), a 2-60 keV
bandpass and a total effective area of 8000~cm$^2$ at 6 keV. The HEXTE
consists of two clusters of four detectors with a total area of
1600~cm$^2$ in the 10 to 250 keV band. As the PCA, it is a non-imaging
device with the same field of view as the PCA. The 1 ks exposure time
is insufficient for a HEXTE detection that allows for a meaningful
analysis. Therefore, we refrain from doing such an analysis. In total
78 observations were carried out with a total PCA exposure of
56~ks. \citet{jonker07} discussed the first 55 observations.

The PCA monitoring campaign also provided triggers for longer
target-of-opportunity observations (TOOs) with the same instrument
(under observation id 92020) once the flux rose to a higher value. We
carried out three TOO campaigns, for a total exposure of 48 ks.

\subsection{{\it Swift}}
\label{sectionswift}

The {\it Swift} mission, launched in September 2004, is an observatory
with a quick slewing time (of order 100~s), which is primarily used to
study gamma-ray bursts (GRBs). It carries three instruments. The Burst
Alert Telescope \citep[BAT;][]{bar05} has a wide field of view (2 sr)
in a 15-150 keV bandpass, with full-width at half-maximum resolutions
of 17\arcmin\ and 7 keV, respectively. The large field of view results
in many serendipitous measurements of hard X-ray sources
\citep{kri06}, including \bron\ (see Fig.~\ref{rxte} middle
panel). The BAT detects randomly occurring GRBs in its field of
view. The position of these GRBs is determined automatically on board
and, depending on a figure of merit, the platform is slewed to that
GRB so that it comes in the field of view of the narrow-field X-ray
Telescope \citep[XRT;][]{bur05} and UV Optical Telescope
\citep[UVOT;][]{rom05}. The bandpass and effective area of the BAT is
such that it is also able to detect bright X-ray bursts, which
typically have black body spectra with peak temperatures of
k$T\sim$3~keV. Usually, X-ray bursts do not invoke an automatic slew,
because there is a black-out list of known X-ray bursters on board to
prevent such slews (X-ray bursts do not belong to the primary
scientific objectives of the mission).

\begin{figure}
  \includegraphics[angle=270,width=\columnwidth]{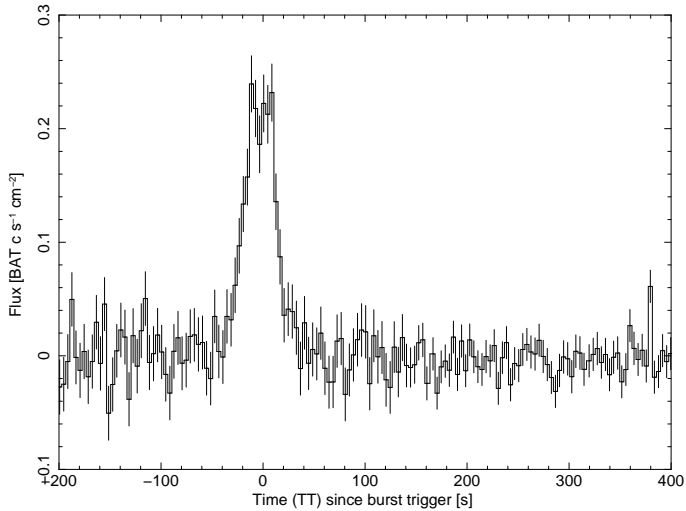}
  \caption[]{Light curve of \bron\ in 15-25 keV data from {\it
      Swift}-BAT, at 4 s resolution. This is a 'mask-weighted' count
    rate history, meaning that the background is subtracted. This
    signal triggered {\it Swift} to slew the narrow-field instruments
    to the source, resulting in the measurement depicted in
    Fig.~\ref{swiftlc1}.
  \label{figbat}}
\end{figure}

On Aug 11, 2006, at 02:59:55 UT, an X-ray burst was detected with BAT
from \bron, which delivered a trigger (trigger number 223918). The BAT
light curve is presented in Fig.~\ref{figbat}. In 15-25 keV, the burst
was detectable for approximately 100 s.  Since at the time \bron\ was
not in the black-out list, this triggered an automatic slew of {\it
Swift} to the 3\arcmin-accurate BAT-determined position. The XRT and
the UVOT were on target after 193~s at 03:03:08 UT. The X-ray burst
was still ongoing and was detected with the XRT
\citep{rom06,kon06}. No signal was detected in the UVOT. {\it Swift}
remained on target for the next 8.76~hr, during 6 orbits, until
11:45:43 UT.

The XRT has a Wolter-I-type mirror system with a focal length of 3.5~m
and a spatial resolution of 18\arcsec\ half-power diameter at 1.5
keV. At its focal plane is a 600$\times$600-pixels CCD with the same
design as the EPIC MOS on XMM-Newton (see Sect. \ref{sectionxmm}). The
field of view captured with the CCD is 23\arcmin$\times24$\arcmin\ and
the pixel size is 2\farcs 36. The peak effective area is 110 cm$^2$ at
1.5 keV. The spectral resolution varies between 50 eV at 0.1 keV to
190 eV at 10 keV (full width at half maximum FWHM).

\begin{figure}
  \includegraphics[angle=270,width=\columnwidth]{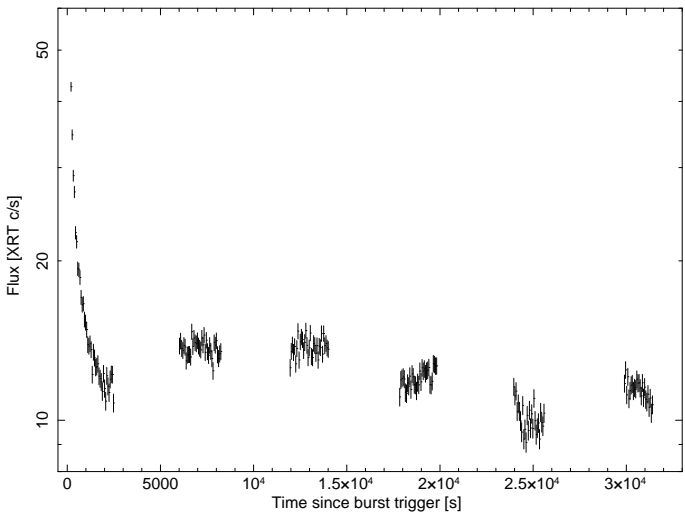}
  \caption[]{Light curve of \bron\ in full bandpass of {\it Swift}-XRT
    at 60 s resolution. The measurements starts 193 s after the burst
    trigger.
  \label{swiftlc1}}
\end{figure}

After acquiring an image for a refined position, whose 4\farcs7-radius
error circle (90\% confidence) contains the optical counterpart of
\bron\ \citep{rom06}, XRT was set to Windowed Timing (WT) mode at
03:08:16 UT for the remainder of the observation. This implies that
the CCD columns are collapsed and only the central 200 out of the 600
pixels are read out, resulting in a 1-dimensional image with a frame
time of 1.7 ms. For the photon count rates encountered in our
observation (50 s$^{-1}$ at maximum), this implies a pile-up fraction
of less than 1\%, which is ignored in the analysis. We employed the
1-dimensional image for the whole observation in sky coordinates to
define an extraction region with a commonly used width of 40\arcsec,
centered on the source, to extract source photons and on an empty part
well outside the point-spread function (PSF) of the source to extract
background photons. The photon rate in the whole bandpass of the
background region was found to be 0.16~s$^{-1}$ and of the source
region at least 10~s$^{-1}$. Thus, the background contribution is
negligible, although it was taken into account throughout our
analysis.

Figure~\ref{swiftlc1} shows a light curve of the photon rate in the
source region of all WT data. The decay of the burst is clearly
visible in the first orbit. The five subsequent orbits show a variable
accretion photon flux between 9 and 14~s$^{-1}$. The burst appears to
decay to a level similar to the average flux in the last orbit (11.5
s$^{-1}$).

The extraction of {\it Swift-}XRT data was done with version 2.0.1 of
the XRT, as embedded in version 2.6.1 of the {\it Swift} software
package. We calculated exposure maps per orbit (taking into account
'hot' pixels) using the tool {\tt xrtexpomap} and effective area
arrays as a function of photon energy per orbit using the tool {\tt
xrtmkarf}.  Spectral bins were grouped such that each resulting bin
contains at least 20 photons to justify the use of the $\chi^2$
statistic as a goodness-of-fit parameter.

{\it Swift}-BAT detected a 2nd X-ray burst on October 13th, 2007 (see
Fig.~\ref{rxte}) that did not trigger XRT follow-up observations.

\subsection{XMM-Newton}
\label{sectionxmm}

XMM-Newton carries 5 X-ray instruments behind 3 identical X-ray
telescopes, each with a 1500 cm$^2$ collecting area. It also carries
the Optical Monitor. For two X-ray telescopes, half of the radiation
goes to two European Photon Imaging Cameras (EPIC) of the Metal-Oxide
Semiconductor (MOS) conductor variety \citep{tur01}. These are sets of
7 front-illuminated CCDs. The other half of the light from the two
telescopes goes to Reflection Grating Spectrometers \citep[RGS1 and
2;][]{her01}. The full light of the third telescope goes to an EPIC
p-n-junction ('pn') CCD, a set of 12 back-illuminated CCDs
\citep{stru01}. The optical monitor (OM) is a 30 cm optical/UV
telescope with a resolution of 1\arcsec, a field of view of 17\arcmin,
a limiting B-magnitude of 20.7 (for 1 ks integration) and a bandpass
of 160 to 600 nm \citep{mason01}.

XMM-Newton observed \bron\ for 41 ks starting Aug 31, 2006, 21:02
UT. EPIC PN was set in Fast Timing mode in which only the central CCD,
covering 13\farcm6$\times$4\farcm4, is read out every 0.03 ms with
preservation of image information along one axis only (the 4\farcm4
side) and full energy information. EPIC MOS1 was set in Full Frame
(imaging) mode while EPIC MOS2 was set in Fast Uncompressed Timing
mode, which is also full imaging mode except for the central CCD that
is read out like EPIC pn in Timing Mode with a time resolution of 1.5
ms. The two EPIC MOS cameras had the filter positioned in 'thin-A'
position and EPIC PN in 'thin'. The MOS1 data suffer somewhat of
out-of-time events. The source position measured from the MOS1 image
is $\alpha_{2000.0}=12^{\rm h}49^{\rm m}39\fs56$,
$\delta_{2000.0}=-59\degr 05\arcmin 18\farcs4$ with an uncertainty of
2\arcsec\ (rms, including systematics). This is 4\farcs0 from the
optical counterpart proposed by \citet{bjzv06}, which is within
2$\sigma$ consistent.  The optical monitor was operated in the 'Image
+ Fast Mode' with the U-band filter in front with a bandpass between
300 and 400 nm.

All data were analyzed with SAS version 7.1.0, and
the standard pipeline products were employed as a basis for extraction
of science data products (light curves and spectra). 

The X-ray observation was hindered by particle flaring for 17\% of the
time according to standard criteria (e.g. for EPIC PN: the rate of
{\tt PATTERN=0} photons above PI channel 10000 was 0.35 s$^{-1}$ or
larger) and this portion of the observation was excluded from further
analysis. For further EPIC pn Timing Mode analysis source photons were
extracted between {\tt RAWX} values of 30 and 46, for pixel patterns
below 5 and grade 0, and for background photons between 10 and 26. The
net source photon rate is $38.08\pm0.04$~s$^{-1}$ and the background
rate 5.3 s$^{-1}$ (full bandpass). For EPIC MOS2 Timing Mode analysis,
photons were extracted with {\tt RAWX} from 294 to 314 and {\tt
PATTERN=0}. Background photons were extracted from the same CCD for
{\tt RAWX} from 258 to 277, which is well outside the PSF of the
source.  The net photon count rate is $10.5\pm0.02$~s$^{-1}$.  The
RGS1 net photon count rate is $0.846\pm0.005$~s$^{-1}$ (1st order
only); for RGS2 this is $1.123\pm0.006$~s$^{-1}$. The light curve
shows no long-term trends during the observation and the Fourier power
spectrum shows white noise above 10 Hz \citep[with a few instrumental
features; ][]{kuster02} and red noise below that. For details on
high-frequency variability see \citet{jonker07}.

We used standard analysis threads for the extraction of X-ray spectra
from EPIC PN and EPIC MOS2 data and employed {\tt rgsproc} for
RGS. The spectral channels were grouped such that each group contains
at least 25 photons. The bandpass for EPIC pn was restricted to
between 0.6 and 12 keV, for MOS2 between 0.3 and 10 keV (following
Kirch 2006) and for RGS between 0.6 and 1.6 keV. A more elaborate
discussion of the RGS spectrum is presented elsewhere (Paerels et al.,
in prep.).

\subsection{Very Large Telescope}

Optical spectra of \bron\ were obtained with FORS2, the low dispersion
spectrograph of ESO's Very Large Telescope in Chile. On April 29,
2006, 2600\,s exposures were taken with the blue 600~lines\,mm$^{-1}$
(600B) and the red 600~lines\,mm$^{-1}$ (600RI) grisms, providing a
mean dispersion of 1.50\,\AA\,pix$^{-1}$ and 1.63\,\AA\,pix$^{-1}$,
respectively. During the blue exposure the seeing was $0\farcs7$, but
it had degraded to $1\farcs3$ during the red exposure, contaminating
the spectrum of the optical counterpart to \bron\ with that of a
nearby star. Two more 2600\,s exposures using the 600RI grism were
obtained on May 5, 2006, but also lacked adequate seeing ($1\farcs3$
and $1\farcs6$). On January 27, 2007 a 2600\,s 600RI spectrum was
obtained under $0\farcs7$ seeing conditions, which fulfilled our
observational constraints. The signal-to-noise of this spectrum is
much better than that of the other three 600RI spectra and in the
analysis below only the 2007 600RI spectrum is used. All observations
used slits of $1\arcsec$ and were obtained in multi-object mode,
capturing spectra of both object X, the counterpart of \bron, and
object A, a nearby blue object \citep{bjzv06}.

The observations were corrected for bias and the sky was subtracted
using clean regions between the stars along the slitlets. The spectra
were extracted using an optimal extraction method similar to that of
Horne et al. (1986) and wavelength and flux calibrated using
observations that were obtained as part of the VLT standard
calibration plan.

\section{Data analysis}
\label{results}

\subsection{X-ray variability}

\begin{figure}
  \includegraphics[width=\columnwidth]{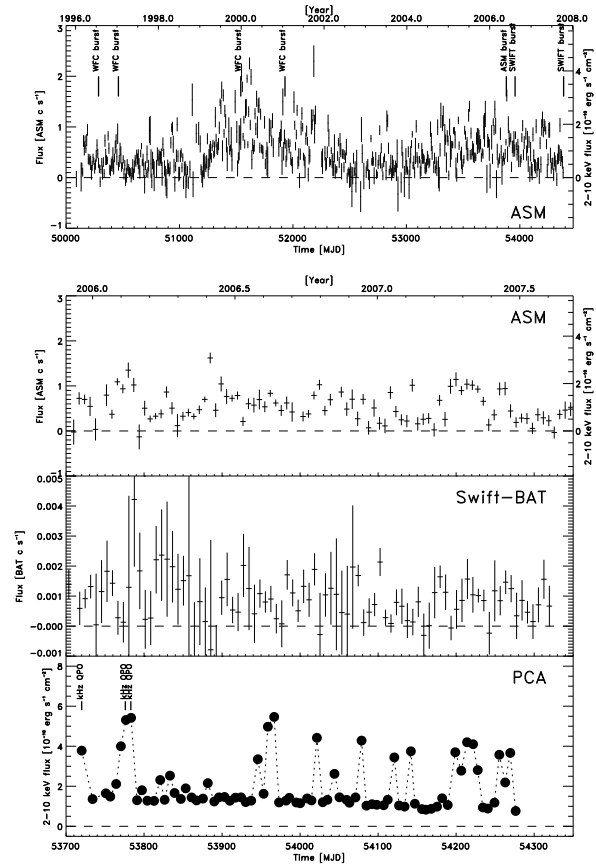} \caption[]{{\it
  (1st panel)\/} 12 year long history of detected photon rate of
  \bron\ as measured with the RXTE ASM. Indicated are the burst
  detections with the BeppoSAX WFCs, ASM and {\it Swift}; {\it (2nd
  panel)} same light curve, zoomed in on last 1.5 yr; {\it (3rd
  panel)} {\it Swift-}BAT-measured history of 15--25 keV photon count
  rate; {\it (4th panel)} PCA-measured history of 2--10 keV energy
  flux. All data have a 1-week time resolution. The last panel
  indicates the data point for which kilohertz QPOs were detected. It
  is an updated version of a similar figure in \citet{jonker07}.
  \label{rxte}}
\end{figure}

In Fig.~\ref{rxte} the monitoring results obtained with ASM, PCA and
BAT are plotted. The ASM light curve (top panel) is the longest
observation series, extending over 12 years. Variability is apparent
on two time scales: one week and hundreds of days. The peak-to-peak
amplitude is a factor of about 8. The average flux is
0.496$\pm$0.005~ASM c~s$^{-1}$, after accounting for a bias level of
0.075 ASM c~s$^{-1}$ \citep{rem97} or roughly
9.4$\times10^{-11}$~\ecs\ (2-10 keV). For a distance of 4.3~kpc (see
Sect.~\ref{bursts}) this translates to a luminosity of
$2.1\times10^{35}$~\lum.

The middle panel shows the BAT monitoring results. These data are of
limited statistical quality, due to the high value of the low-energy
threshold (15 keV), but they show that no unexpected high-energy
behavior occurs.

The bottom panel shows the results of the PCA monitoring campaign
(only obsid 90042-02) in terms of calibrated energy flux.  Average
spectra were extracted for each PCA observation. These were, between 3
and 20 keV, modeled with a comptonized spectrum multiplied by
absorption, see Sect.~\ref{xrayspectrum}.  The equivalent hydrogen
column density was fixed to $N_{\rm H}=4\times10^{21}$~cm$^{-2}$ (see
spectral analysis below), which is hardly noticeable in the 3-20 keV
bandpass. Subsequently, we determined the energy flux between 2 and 10
keV. These data are of much higher quality than the ASM data and they
provide a clearer picture of the variability. It is noted that no
other high-frequency QPOs were found apart from the ones reported by
\citet{jonker07}.

\subsection{X-ray bursts}
\label{bursts}

\begin{figure*}[t]
\begin{center}
\includegraphics[width=0.7\columnwidth]{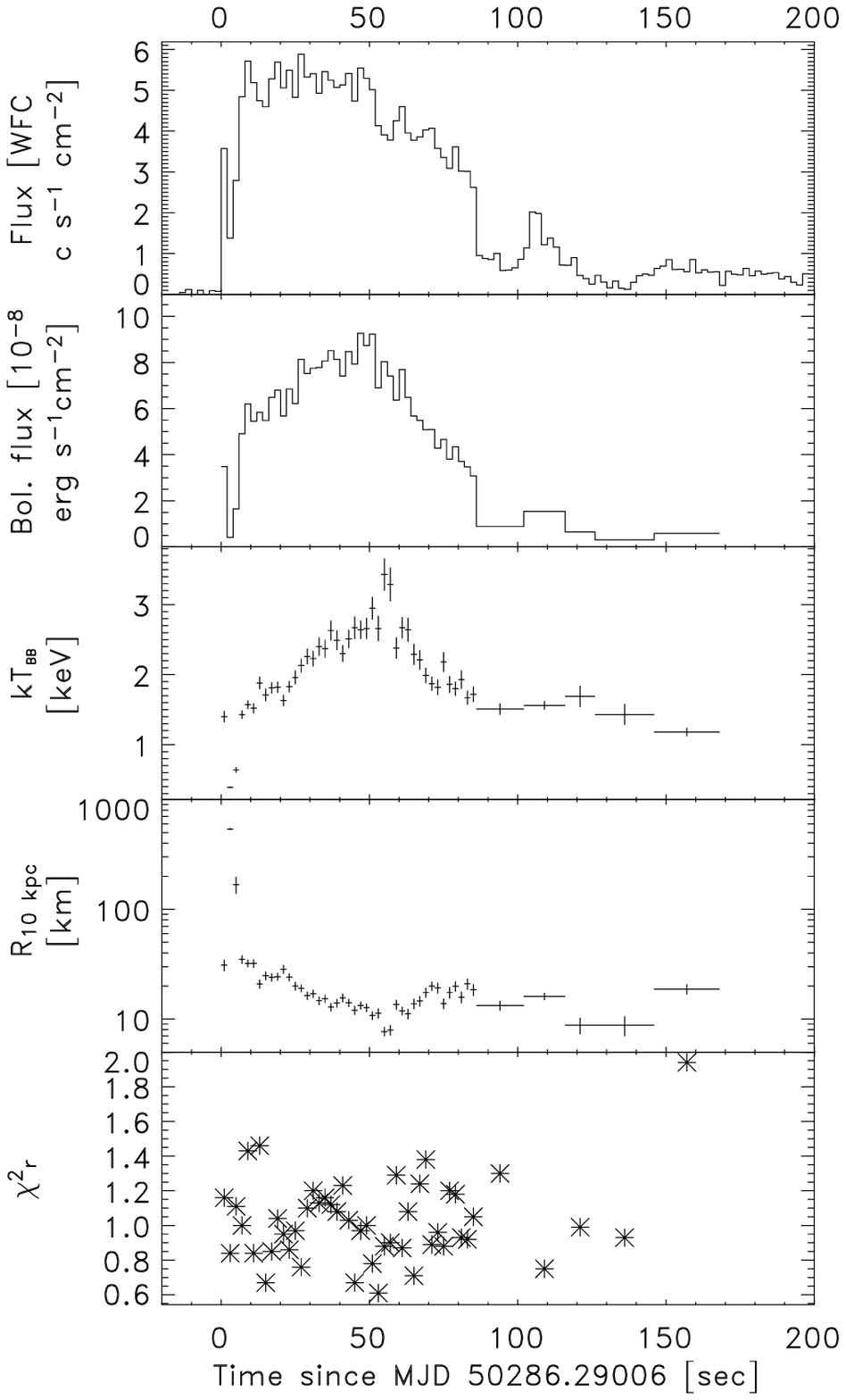}
\includegraphics[width=0.7\columnwidth]{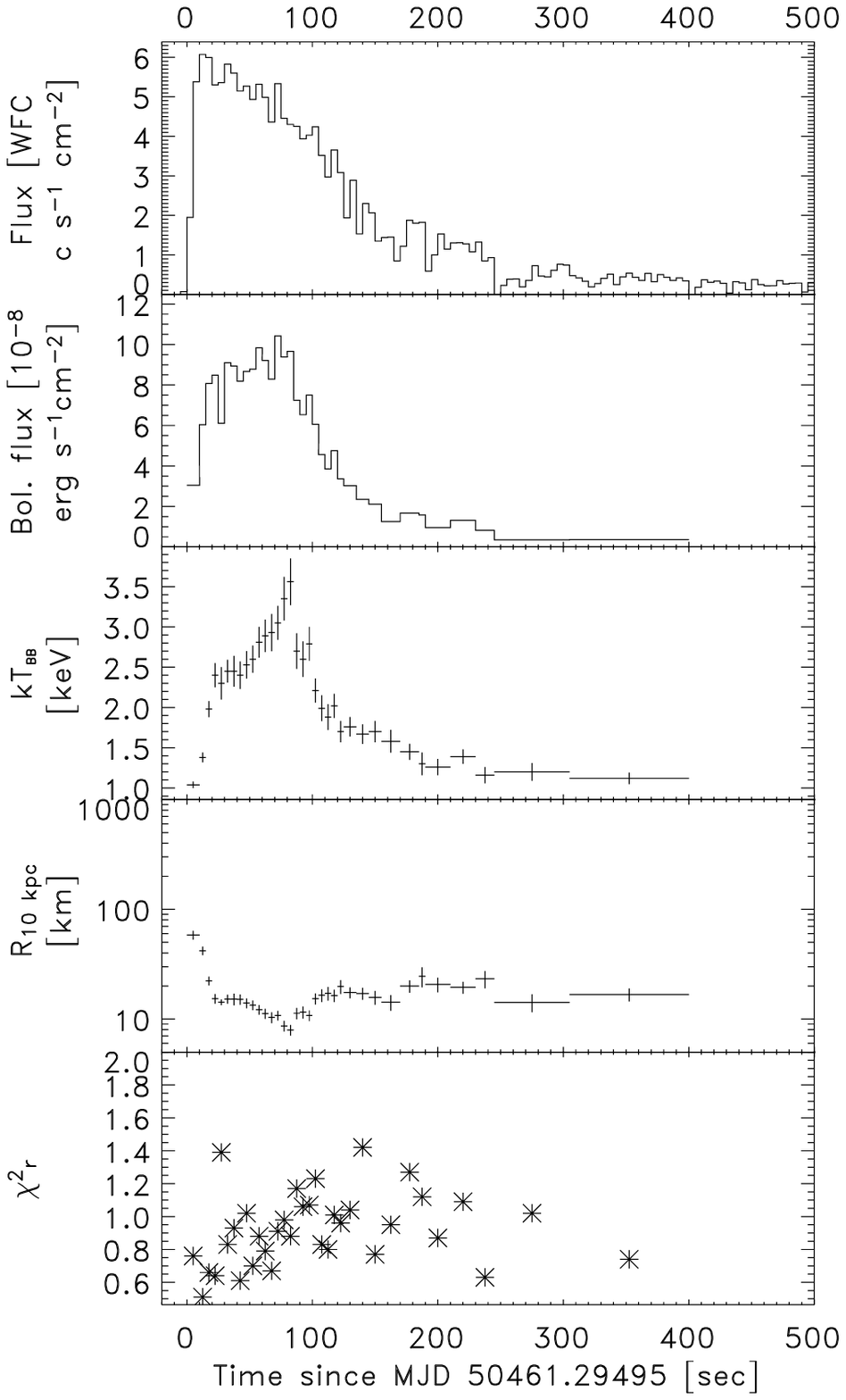}\\
\includegraphics[width=0.7\columnwidth]{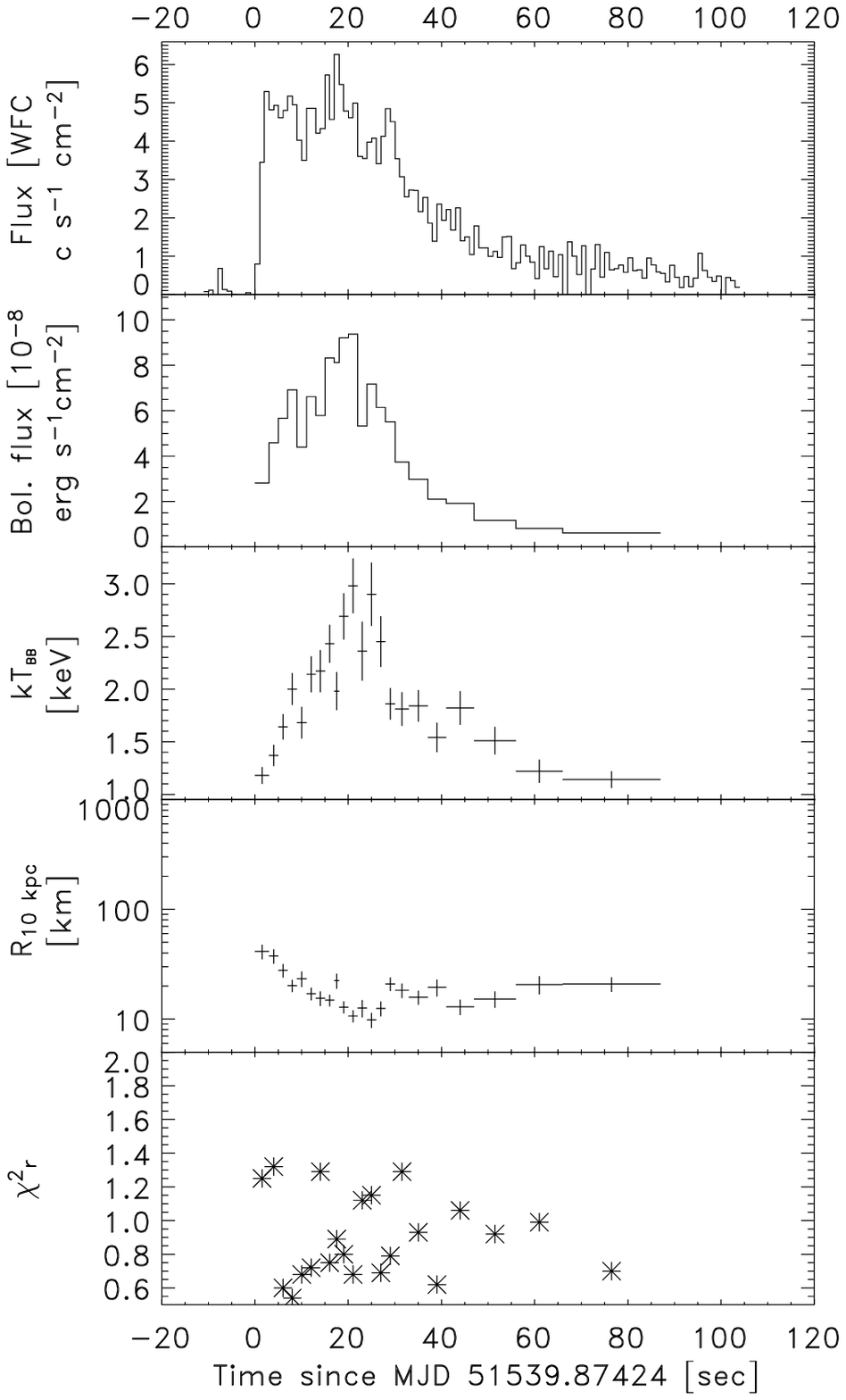}
\includegraphics[width=0.7\columnwidth]{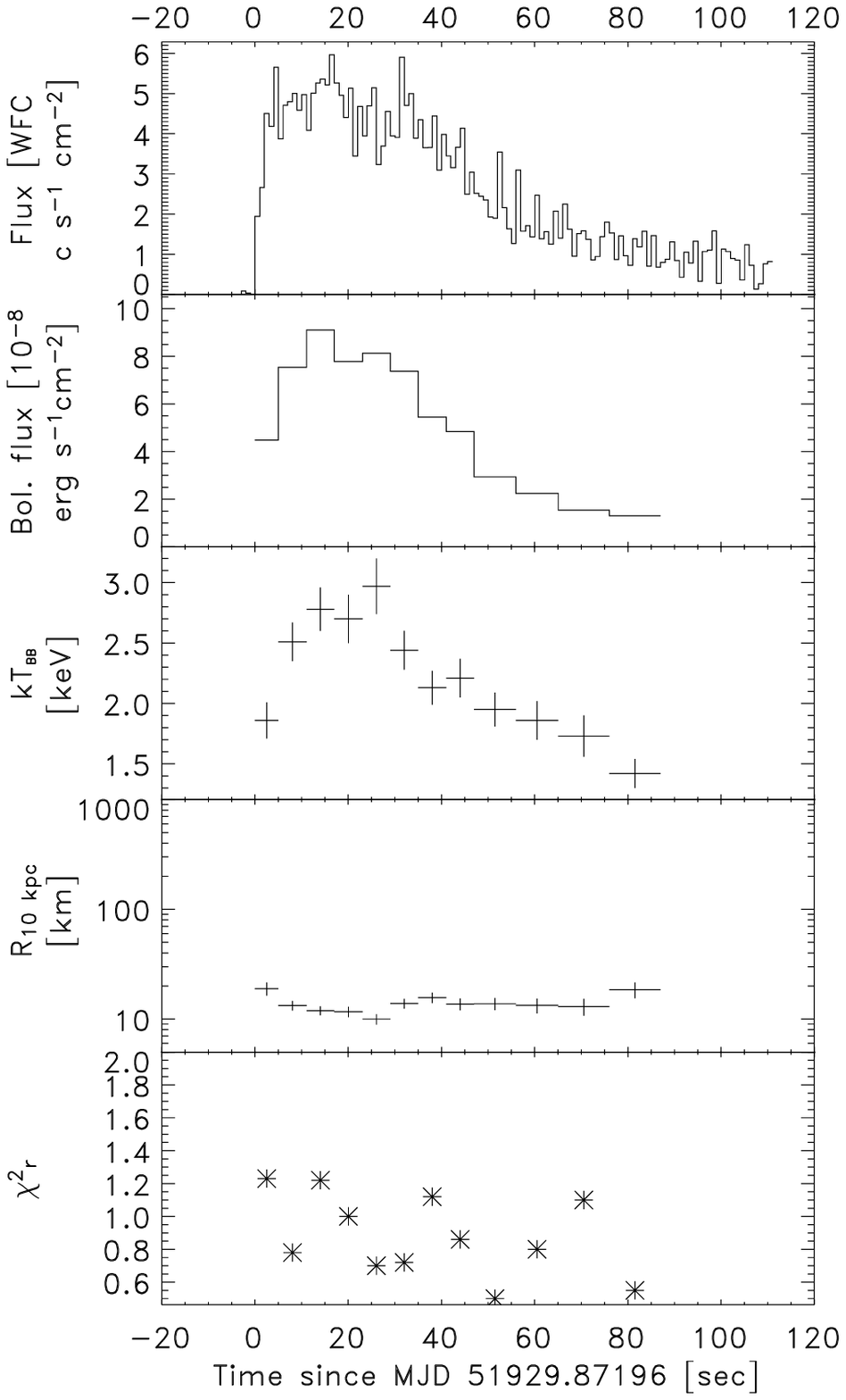}
\caption{Time-resolved spectroscopy of the 4 bursts detected from
\bron\ with the BeppoSAX WFCs. \label{wfcbursts}}
\end{center}
\end{figure*}

\begin{figure}
  \centering
  \includegraphics[width=\columnwidth]{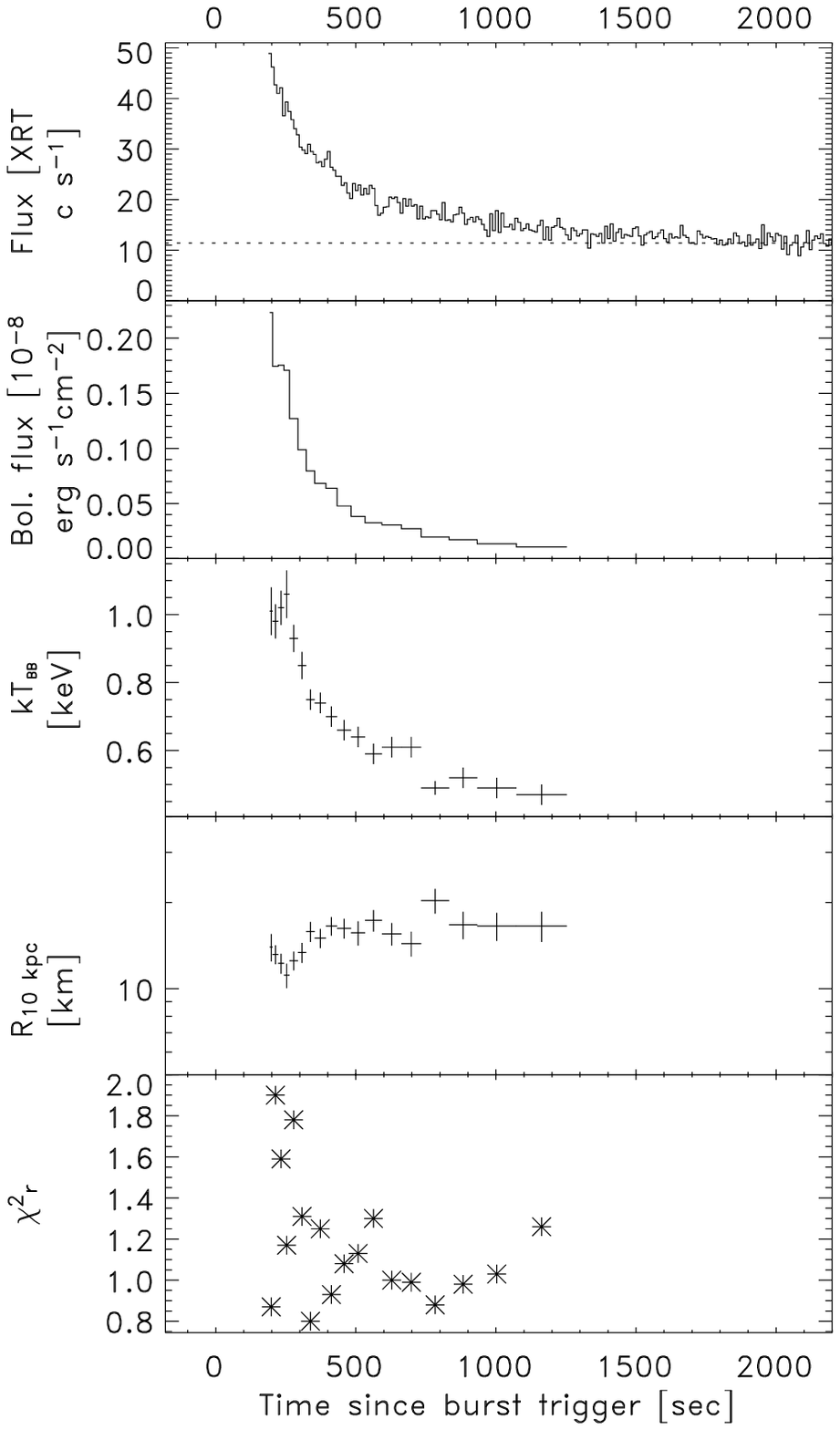}
  \caption{Time-resolved spectroscopy of X-ray burst detected with
    {\it Swift}-XRT. Note the lack of XRT data during the first 201 s
    of the burst. Therefore, the brightest parts of the burst were
    missed by this instrument.\label{swiftburst}}
\end{figure}

We carried out a time-resolved spectral analysis\footnote{{\tt XSPEC}
version 12.3.1x \citep[][]{arn96,dor01} was employed for X-ray
spectral analyses.} of the four WFC bursts by modeling the spectrum
with an absorbed black body (for details about the absorption model,
see Sect. \ref{xrayspectrum}) as a function of time and measuring the
evolution of temperature and emission area, see
Fig.~\ref{wfcbursts}. The persistent emission was neglected because it
is less than 1\% of the total emission at the burst peak.

All bursts are fairly long, with e-folding decay times between about
50 and 100~s. Also, they are Eddington limited. The effect of the
Eddington limit is most clearly seen in the evolution of the size of
the emission region for the first two bursts as measured through the
radius of an isotropically radiating sphere. There is clear expansion
of the photosphere. In particular, during the first burst the
photosphere expands by two orders of magnitude in radius. In cases
where a change in emission region size is not strong, an indication of
near-Eddington fluxes is suggested by the prolonged rising of the
temperature. This takes up to 100 s to reach the peak value for the
first two WFC bursts and 20 to 30 s for the latter two. The
long-duration rise of the temperature is likely due to a photosphere
that is receeding to the neutron star surface after an expansion due
to super-Eddington fluxes. In non-Eddington-limited bursts, rising
high-energy fluxes last less than 1~s unless hydrogen is present, which
in our ultracompact case is unlikely. The peak unabsorbed bolometric
flux is comparable over the four WFC bursts, at
$(9.5\pm0.5)\times10^{-8}$~\ecs. If this is equivalent to the
Eddington limit for a hydrogen-poor atmosphere (as expected for a
UCXB) of $3.8\times10^{38}$~\lum, we derive a distance of
5.8~kpc. Correcting for a gravitational redshift of $z=0.35$ at the
surface of a canonical neutron star (i.e., with a mass of
1.4~M$_\odot$ and a radius of 10 km) decreases the luminosity for a
distant observer by a factor of $(1+z)^{-2}=0.55$
\cite[e.g.,][]{lew93}. As a result, the distance decreases by a factor
of 0.75 to 4.3 kpc. Furthermore, scattering of the black body photons
by electrons in the NS atmosphere, particularly when the flux is near
Eddington, will alter the spectrum and at the least introduce
uncertainties in the distance of order 10\% \citep[e.g.,][]{lon86}.

The {\it Swift} burst was initially detected with the BAT and followed
up with the XRT. One spectrum was extracted from the BAT data, during
the peak between -20.0 to +9.7 s from the trigger time (see
Fig.~\ref{figbat}). Applying a higher time resolution is not useful
given the low count rate.  The 15-30 keV spectrum was fitted with a
black body (absorption is not relevant in this bandpass) and we find
k$T=2.8\pm0.2$~keV and a flux of
$(1.4\pm0.2)\times10^{-8}$~\ecs\ ($\chi^2_\nu=0.55$ for
$\nu=10$). This translates to a bolometric flux of
$7.4\times10^{-8}$~\ecs, which is, within error margins, equal to the
value found for the WFC bursts. The longevity of the rising $>15$ keV
flux (35 s) suggests super-Eddington fluxes.

The data from the sensitive XRT traces the faint end of a burst and it
is necessary to take into account the persistent emission: the
persistent spectrum was modeled by a combination of a power law and a
black body multiplied by absorption, fixing the parameters for the
absorption and power law to that found for the 6th orbit whose photon
rate is closest to the aymptotic value found during orbit 1.  The
results for the time-resolved spectroscopy of the burst in XRT are
presented in Fig.~\ref{swiftburst}. The XRT data provide a rare
opportunity to study a burst down to relatively low temperatures. This
opportunity is facilitated by the combination of a burster with bright
bursts, low persistent flux and low absorption, and an instrument that
(compared to instruments that mostly observe bursts) is sensitive, has
low background levels and covers sub-keV photon energies. It is
possible to follow burst temperatures down to less than 0.5~keV and up
to half an hour after the onset. There are no unexpected measurements
for this burst, except for the long duration. The spectrum shows no
narrow features for any choice of time interval, and the decay is as
expected for a UCXB with a low accretion rate \citep[e.g.,][]{zan07}.

An interesting aspect is the occurrence of irregularities from a
monotonic decay for the first two WFC bursts. This is similar to what
is seen in some other long bursts from UCXBs \citep{zan05a,zan07} and
is not explained. For certain, \bron\ is not a dipping source as dips
should have been detectable in the persistent emission.

From the 8 Msec WFC exposure time, the average burst recurrence time
is very long: $23\pm12$~d, which is consistent with a low-mass
accretion rate \citep[for details, see][]{zan07}. The closest observed
burst pair had a 77 d recurrence time \citep{lev06,rom06}.

We tested the BAT data for the occurrence of (quasi) periodic signals,
motivated by the positive detection with BAT of a 414~Hz burst
oscillation in the UCXB 4U 0614+091 \citep{stroh08}, although we
realize that chances for detection are small because of the six times
smaller raw photon count rate during the burst peak. We selected BAT
detectors that had $>50$\% exposure by \bron. Such a selection
improves the sensitivity through a decrease of the background by a
factor of 3 at a mere 35\% loss of source photon rate. The time
resolution is 100~$\mu$s. We searched the 13-25 keV photon rate with
standard Fourier techniques \citep[e.g.,][]{stroh08} and found no
oscillations. The 2-sigma upper limit for a 1 Hz wide QPO is 8\%
r.m.s.

On May 26, 2006, a type-I X-ray burst was detected with the ASM
\citep{lev06} lasting at least 340 s (there were 4 consequetive dwells
on the source) and a peak flux of 1.91 Crab units averaged over the
second dwell, which is equivalent to approximately
$4\times10^{-8}$~\ecs\ (2-12 keV). The peak temperature of 2.8 keV
\citep{kuu06} is consistent with the measurements of the other bursts.

\subsection{X-ray spectrum of persistent emission}
\label{xrayspectrum}

\begin{figure}[!t]
  \includegraphics[angle=270,width=\columnwidth]{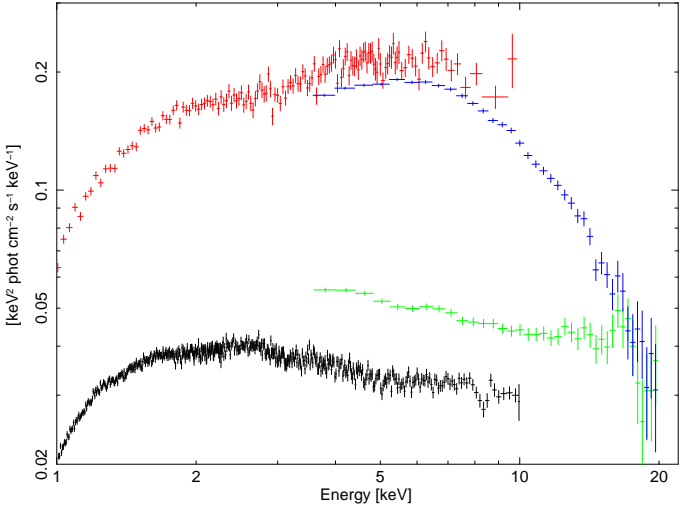}
  \caption[]{Calibrated-flux spectra of two PCA observations (green
    and blue data points, covering 3-20 keV), XMM-Newton/EPIC pn
    (black; for illustrative purposes only EPIC pn data are shown) and
    {\it Swift-}XRT (red). We choose to exclude sub-1 keV data from
    this plot to focus on the parts where the spectral changes are
    largest.
  \label{fig4spectra}}
\end{figure}

We first concentrate on a subset of four observations that trace the
flux range well and have good-quality spectra. These are the
XMM-Newton, the {\it Swift-}XRT and two long PCA measurements at low
and high fluxes (obsids 92020-01-04-00 [exposure time 10752 s] and
92020-01-05-00 [8192 s] respectively). Figure~\ref{fig4spectra} shows
these spectra in calibrated flux units.  It shows that only in the
high-flux PCA observation curvature is detected that is not related to
the low-energy absorption. Specifically, there is a high-energy
cutoff. The three other spectra are fairly well represented by
(absorbed) power laws, see Table~\ref{tabfit1}.

Curvature may be introduced in the broad-band spectrum of the
persistent emission from an X-ray burster through a generic spectral
continuum model \citep[e.g.,][]{whi88,mit89,sid01} that is a
combination of a 'disk black body' \citep[model {\tt diskbb} in
XSPEC;][]{mit84,maki86} and comptonization of soft photons by a hot
plasma \citep[model {\tt comptt} in XSPEC;][]{tit94}. If the upper
boundary of the bandpass is limited to 10 or 20~keV, it is difficult
to constrain the comptonized component since the plasma temperatures
are often higher than that boundary. If the optical depth is low, the
comptonized component will, within the bandpass, simplify to a power
law. Disk black body temperatures are generally lower than the plasma
temperatures; the interpretation favors the above model only if the
spectrum is concave pointing up. If it is concave pointing down (i.e.,
there is a high-energy cutoff), the break is more likely due to a
detection of the plasma temperature. In our case, if curvature is
detected it is always in the latter sense and the evidence for the
presence of a disk black body component is not significant. A model is
favored with just a single continuum component consisting of a
comptonized component. All spectra are indeed consistent with this
model.

\begin{table}
\caption[]{Power-law fits to 4 spectra in
  Fig.~\ref{fig4spectra}\label{tabfit1}. The high $\chi^2_\nu$ value
  for the 'PCA high' spectrum points to the inconsistency with a
  power-law model.}
\begin{center}
\begin{tabular}{lllll}
\hline\hline
Parameter & PCA low & PCA high & {\it Swift}-XRT & XMM-Newton \\
\hline
Date & 2006-08-21 & 2007-04-10 & 2006-08-11 & 2006-08-31 \\
Bandpass  & 3-20~keV & 3-20 & 0.3-10 & 0.6-10 \\
$N_{\rm H}^\ast$ & 0.51 (fixed) & 0.51 (fixed) & $0.459\pm0.09$ & $0.577\pm0.002$ \\
$Z_{\rm Ne}$ & n/a & n/a & $3.3\pm0.4$ & $1.89\pm0.06$ \\
$\Gamma$ & $2.27\pm0.02$ & 2.48 & $1.93\pm0.03$ & $2.348\pm0.006$ \\
Flux$^\ddag$ & $1.37\pm0.01$ & 4.85 & $5.10\pm0.05$ & $0.887\pm0.003$ \\
$\chi^2_\nu$ ($\nu$) & 0.8851 (37) & 91.0$^\times$ (37) & 1.209 (219) & 1.3451 (1891) \\
\hline\hline
\end{tabular}
\end{center}

$^\ast$in units of $10^{22}$~cm$^{-2}$; $^\ddag$2-10 keV absorbed
flux, in 10$^{-10}$~\ecs; $^\times$The reduced $\chi^2$ is so high
that we refrain from quoting errors on fitted parameters.

\end{table}

\begin{figure}[!t]
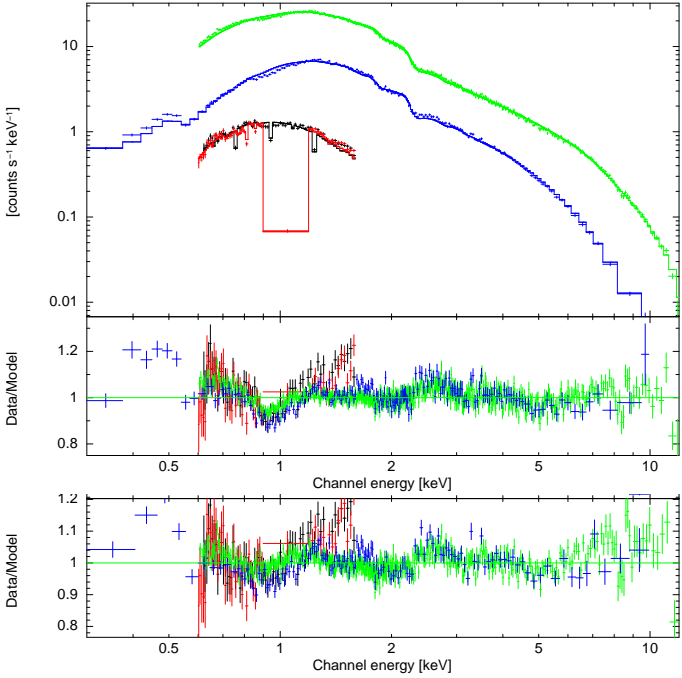

  \includegraphics[angle=270,width=\columnwidth]{9361f7a.ps}\\
  \includegraphics[angle=270,width=\columnwidth,clip]{9361f7b.ps}
  \caption[]{Spectrum of \bron\ as measured with RGS1 (red), RGS2
    (black), EPIC PN (green) and MOS2 (blue). The upper panel shows
    the spectra with the fitted comptonized model with a Neon
    abundance fixed to the solar value, the middle panel the residuals
    in terms of model ratio and the lower panel the residuals after
    leaving free the Neon abundance.
  \label{xmmspectrum}}
\end{figure}

\begin{figure}[!t]
  \includegraphics[angle=0,width=\columnwidth]{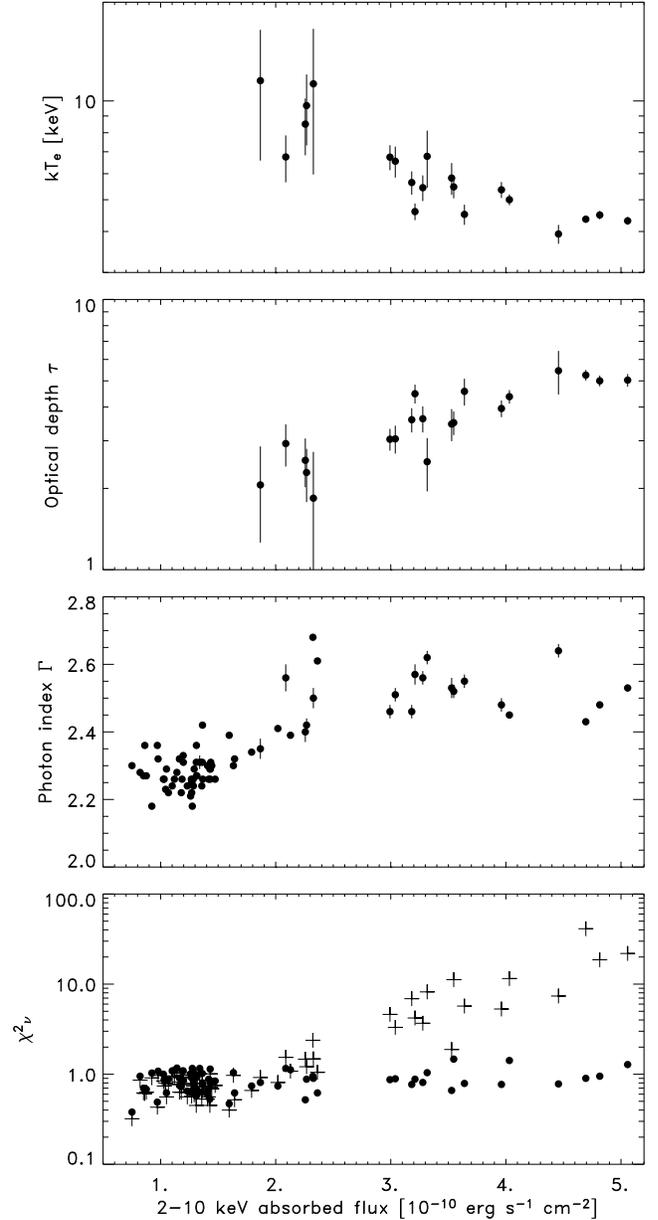}
  \caption[]{(Upper two panels) plasma temperature and optical depth
    for a fit of a comptonized spectrum to the 3-20 keV PCA monitoring
    data as a function of 2--10 keV (absorbed) flux. (3rd panel)
    photon index for a power-law fit. (last panel) Goodness-of-fit
    values for power law (crosses) and comptonized fits (filled
    circles). Errors are single-parameter $1\sigma$ values and
    underestimate the true error when $\chi^2_\nu$ is far from
    acceptable (for the power-law fits for fluxes above
    $2\times10^{-10}$~\ecs).
  \label{pcaspectra}}
\end{figure}

\begin{table}
\caption[]{Best-fit parameters for single component spectra.
A disk geometry is assumed for the comptonized spectra.\label{tabfit2}}
\begin{center}
\begin{tabular}{ll}
\hline
\multicolumn{2}{c}{\it Swift-XRT+PCA(high) / comptonized} \\
\hline
$N_{\rm H}$ (10$^{21}$~cm$^{-2}$) & $2.1\pm0.3$ \\
$Z_{\rm Ne}$                     & $5.9\pm0.9$ \\
k$T_0$ (keV)                & $0.26\pm0.02$ \\
k$T_{\rm e}$ (keV)               & $2.39\pm0.04$\\
$\tau$                          &  $7.00\pm0.13$\\
$\chi^2_\nu$ (dof)              & 1.034 (255) \\
Flux$^\ddag$ 2-10 keV (absorbed)        & $4.94\pm0.06$ \\
Flux$^\ddag$ 0.3-10 keV (absorbed)      & $6.70\pm0.07$ \\
Flux$^\ddag$ 0.3-10 keV (unabsorbed)    & 7.72 \\
Flux$^\ddag$ 0.1-100 keV (unabsorbed)   & 8.90 \\
\hline
\multicolumn{2}{c}{\it Swift-XRT+PCA(high) / comptonized + disk black body} \\
\hline
$\chi^2_\nu$ (dof)              & 1.035 (254) \\
\hline
\multicolumn{2}{c}{\it XMM-Newton (RGS+EPIC-PN+MOS2) / comptonized} \\
\hline
$N_{\rm H}$ (10$^{21}$~cm$^{-2}$) & $(2.53\pm0.03)$ \\
$Z_{\rm Ne}$                     & $2.86\pm0.15$ \\
k$T_0$                        & $0.260\pm0.003$ \\
k$T_{\rm e}$                   & 288 \\
$\tau$                        & $0.0190\pm0.0005$ \\
$\chi^2_\nu$ (dof)              & 1.3304 (4657) \\
Flux$^\ddag$ 2-10 keV (absorbed)        & $0.888\pm0.004$ \\
Flux$^\ddag$ 0.3-10 keV (absorbed)      & $1.361\pm0.003$ \\
Flux$^\ddag$ 0.3-10 keV (unabsorbed)    & 1.671 \\
Flux$^\ddag$ 0.1-100 keV (unabsorbed)   & 2.437 \\
\hline
\multicolumn{2}{c}{\it XMM-Newton (RGS+EPIC-PN+MOS2) / power law} \\
\hline
$N_{\rm H}$ (10$^{21}$~cm$^{-2}$) & $(5.78\pm0.03)$ \\
$Z_{\rm Ne}$                     & $1.95\pm0.05$ \\
$\Gamma$                        & $2.359\pm0.005$ \\
$\chi^2_\nu$ (dof)              & 1.398 (4659) \\
Flux$^\ddag$ 2-10 keV (absorbed)        & $0.886\pm0.004$ \\
Flux$^\ddag$ 0.3-10 keV (absorbed)      & $1.356\pm0.003$ \\
Flux$^\ddag$ 0.3-10 keV (unabsorbed)    & 3.015 \\
Flux$^\ddag$ 0.1-100 keV (unabsorbed)   & 5.75 \\
\hline
\multicolumn{2}{c}{\it PCA (low) / power law} \\
\hline
$N_{\rm H}$ (10$^{21}$~cm$^{-2}$) & $<0.3$ \\
$Z_{\rm Ne}$                     & unconstrained \\
$\Gamma$                        & $2.26\pm0.02$ \\
$\chi^2_\nu$ (dof)              & 1.226 (37) \\
Flux$^\ddag$ 2-10 keV (absorbed)        & $0.91\pm0.02$ \\
Flux$^\ddag$ 0.3-10 keV (absorbed)      & $1.356\pm0.003$ \\
Flux$^\ddag$ 0.3-10 keV (unabsorbed)    & 2.63 \\
Flux$^\ddag$ 0.1-100 keV (unabsorbed)   & 4.89 \\

\hline\hline
\end{tabular}

$^\ddag$in 10$^{-10}$~\ecs.
\end{center}
\end{table}

There is significant low-energy absorption in the spectrum. We modeled
this by photo-electric absorption in a cold gas of solar abundances
\citep[following][]{lodd03} and with absorption cross sections
according to \citet{bmc92}. The {\tt vphabs} model in {\tt XSPEC} is
employed that allows for fitting abundances per element in the
absorbing gas. While for most elements this did not result in
significantly better fits, Neon is the exception. For any continuum
model and for data covering sub-keV photon energies, leaving free the
Neon abundance yields better fits.  We specify the best-fit Neon
abundance with respect to solar, $Z_{\rm Ne}$, when appropriate. The
relevance of Neon is visible in Fig.~\ref{xmmspectrum}, which shows
the XMM-Newton spectra fitted with a continuum model without leaving
$Z_{\rm Ne}$ free. The fit shows large negative residuals at 1~keV
that can be explained by the Neon K-shell absorption edge with a
larger-than-solar abundance.

We analyzed the {\it Swift-}XRT and PCA high-flux spectra
simultaneously . Since the fluxes are so similar
(Table~\ref{tabfit1}), as are the spectral shapes in the overlapping
3-10 keV band (Fig.~\ref{fig4spectra}), the spectral shape is assumed
to be the same in both cases. To allow for the slightly different
fluxes, a normalization factor is introduced between both data
sets. The results of the fits are presented in Table~\ref{tabfit2}. As
a test, a disk black body is added and it is found that the fit does
not improve (see Table~\ref{tabfit2}).

The X-ray spectra from XMM-Newton probe the lowest fluxes, which, in a
sense, is fortunate because these are the most sensitive
data. However, the analysis of these spectra is cumbersome. The fit
with a comptonized model is slightly better than with a power law, see
Table~\ref{tabfit2}, with a plasma temperature above the bandpass, but
none of the models provide satisfactory fits. It is suspected that
this is mainly due to calibration uncertainties rather than to
incompleteness of the models. This is most clearly illustrated by the
inconsistencies between the results of pairs of instruments. For
instance, there are clear deviations between the RGS and CCD spectra
and between both CCD spectra above 5 keV. Furthermore, the deviations
between the CCD spectra and the models (see lower panels in
Fig.~\ref{xmmspectrum}) show a jump at 2.3 keV that is coincident with
a jump in the effective area for both instruments. This jump is
related to the telescope mirrors and we refrain from interpreting this
as a true effect in the spectrum of \bron. If the spectra of all
instruments are individually modeled, introduction of systematic
uncertainties of 5, 3, and 1\% can lower $\chi^2_\nu$ to an acceptable
1 for EPIC MOS, PN and RGS, respectively. Thus, the timing-mode MOS2
spectrum seems most in need for a refined calibration. The apparent
uncertainties in the calibration make us hold back from interpreting
the details of the spectra. We only consider relevant the results of
the general continuum model and the high-resolution features seen in
the RGS spectrum.

There are no narrow spectral lines in the high-resolution RGS
spectra. Typical upper limits on the equivalent width of narrow lines
are 3 eV at 1 keV (with coverage by only RGS1) and 2 eV at 1.2 keV
(with coverage by both RGSs).  However, absorption edges are clearly
detected and these high-resolution spectra provide the most accurate
means for abundance determinations. The K-edges of neutral O and Ne
(at 0.538 keV and 0.870 keV, respectively) and the L-edge of neutral
Fe (0.7 keV) are significantly detected. To avoid complications in
modeling these edges due to the various other manifestations of these
elements \citep[ionized species, atoms contained in dust particles;
e.g.,][]{kaa07}, we determined the edge depths by zooming in on them
and ignoring data at the exact energy of the edge. Thus, we only
analyzed the following ranges of photon energy: 0.480-0.524 and
0.550-0.600 keV for the O K-edge; 0.600-0.690 and 0.710-0.800 keV for
the Fe L-edge; and 0.700-0.850 and 0.880-1.100 keV for the Ne
K-edge. The continuum model is fixed to that found from all XMM-Newton
data, however, we left free the power-law normalization (see
Table~\ref{tabfit2}) and the abundances of the three elements in
question, fitting each of the 3 edges separately. The 'Cash' statistic
\citep{cash} was used to find the best fit. The results are: $Z_{\rm
O}=0.974\pm0.041$, $Z_{\rm Fe}=1.449\pm0.097$ and $Z_{\rm
Ne}=1.586\pm0.128$. The value for Ne differs 2.6$\sigma$ from the
combined EPIC/RGS analysis. The Ne/O abundance ratio with respect to
solar \citep[as prescribed by][]{lodd03} is $1.62\pm0.15$, 4$\sigma$
above solar.

We provide the results of modeling the PCA monitoring data in
Fig.~\ref{pcaspectra}. We modeled the data with an absorbed
comptonized spectrum as well as an absorbed power law. This again
shows that the spectra for the higher fluxes are only well fitted with
the comptonized spectrum. The power-law fits fail for 2-10 keV fluxes
above $2\times10^{-10}$~\ecs. In other words, it is above this flux
threshold that the plasma temperature drops below the 20~keV boundary
of the bandpass.

\begin{figure}
  \includegraphics[angle=0,width=\columnwidth]{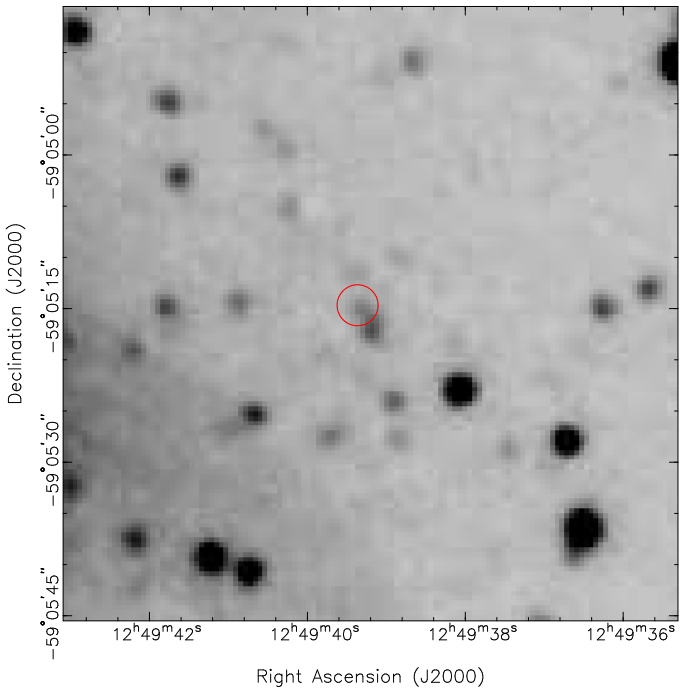}
  \caption[]{This panel shows a 35\arcsec-wide part of U-band image
  from XMM OM. Images per orbit were added after determination of the
  relative shifts (in units of integer pixels) from stars visible in
  each image. The counterpart to \bron\ is encircled
  \citep[c.f.,][]{bjzv06}. The position is calibrated through
  astrometric comparison with archival discovery data of the optical
  counterpart and is accurate within 0\farcs3.  \label{figom}}
\end{figure}

\begin{figure*}
  \centering \includegraphics[width=17cm]{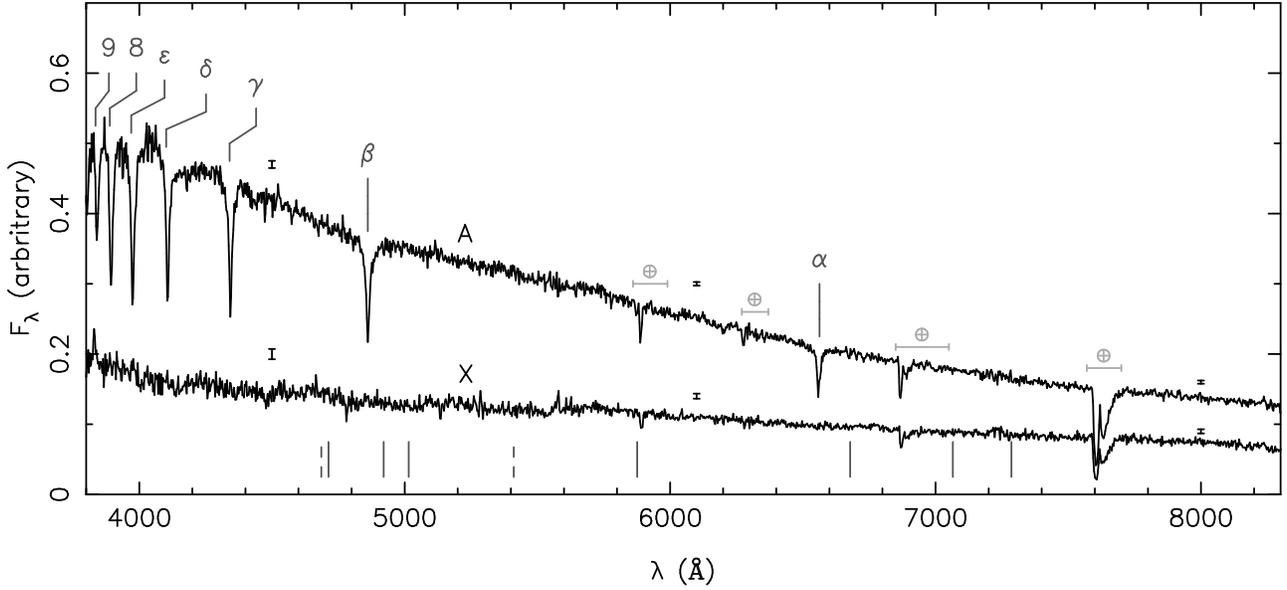}
  \caption{We show the optical spectrum of object X, the optical
  counterpart to \bron\ and object A, and the nearby blue star. Both
  spectra are binned by averaging each consecutive pair of data
  points. Typical uncertainties in the flux of both objects are shown
  at 4500\,\AA, 6100\,\AA\ and 8000\,\AA. The vertical lines above the
  spectrum of object A indicate the wavelengths for the hydrogen
  Balmer lines from H$\alpha$ upto H9. The vertical lines below the
  spectrum of object X indicate the wavelengths at which helium I
  lines (solid) and the helium II line at 4686\,\AA\ (dashed) are
  located. Regions where telluric absorption is present are denoted by
  the horizontal bars above the spectrum of object A.}
  \label{fig:spectrum}
\end{figure*}

\subsection{Optical photometry}

A U-band exposure taken with the XMM-Newton OM (exposure time 36 ks)
is shown in Fig.~\ref{figom}. The optical counterpart to \bron\ is
clearly detected. It is blended with a nearby star, which precludes an
accurate assessment of the U-band brightness. A photon list was
recorded for a 22$\times$23 pixels part of this image, which includes
the optical counterpart, with a time resolution of 0.5~s. We extracted
the photons from a region that includes the counterpart and excludes
as much as possible the blended star. The selected set of 2404 photons
(for an average rate of 0.067 s$^{-1}$) was explored for periodicities
using Fourier power and epoch-folding techniques. No periodicity was
found up to 20$\times10^3$~s, with an upper limit of about 5\% in
amplitude (0.05 in uncalibrated magnitude).

\subsection{Optical spectroscopy}

Figure\,\ref{fig:spectrum} shows the combined VLT spectra of objects A
and X \citep[the two blue objects in the field found
by][]{bjzv06}. The hydrogen Balmer lines H$\alpha$ up to H9 are seen
in the spectrum of star A. Due to the strong pressure broadening it is
clear that the star is a white dwarf. A comparison of its spectrum to
hydrogen atmosphere models by Pierre Bergeron (2003, private
communication) indicates that it is of spectral type DA7 and possibly
has a surface gravity, $\log g\la7$, that is low for a white dwarf.

The spectrum of object X, the optical counterpart to \bron, is
featureless except for absorption features, which can be attributed to
telluric absorption and are also seen in the spectrum of object A. In
order to quantify the presence or absence of spectral features we
follow the method employed by \citet{njmk04}; we determine equivalent
widths (EWs) for different spectral ranges. This is done by fitting a
straight line to the 25\,\AA\ regions of the continuum on either side
of the spectral range in question and this model for the continuum is
subtracted from the data points inside the spectral range. By
computing the weighted average of these datapoints and combining it
with the width of the spectral range we obtain 'area' of the spectral
feature that may be present and use it to define the EW. Using this
strategy, no assumptions about the shape of the spectral feature have
to be made.

\begin{table*}
  \centering
  \caption[]{Negative equivalent width ($-$EW) measurements and upper limits
  ($3\sigma$). Negative $-$EW measurements denote absorption features,
  positive emission. The upper limits are limits on both absorption and
  emission features}\label{tab:ew}
  \renewcommand{\footnoterule}{}
  \begin{tabular}{lccc||lcc|lcc}
    \hline \hline
    Line & $\lambda$ & Spectral range & $-$EW &
    Line & Spectral range & $-$EW &
    Line & Spectral range & $-$EW \\
    & (\AA) & (\AA) & (\AA) & & (\AA) & (\AA) & & (\AA) & (\AA) \\
    \hline
    H$\delta$        & 4101 & 4091--4111 & $<0.98$        & \ion{C}{iii}     & 4624--4680 & $1.73\pm0.44$  & \ion{O}{ii}     & 4680--4720 & $<1.10$\\ 	 
    H$\gamma$        & 4340 & 4330--4350 & $<0.87$        & \ion{C}{ii}      & 5110--5165 & $-2.88\pm0.40$ & \ion{O}{ii}     & 4900--4960 & $<1.33$\\ 
    H$\beta$         & 4861 & 4851--4871 & $<0.77$        & \ion{C}{iii}     & 5230--5310 & $4.81\pm0.52$  & \ion{O}{ii}     & 5160--5220 & $2.25\pm0.44$\\
    H$\alpha$        & 6563 & 6552--6572 & $<0.64$        & \ion{C}{iii}     & 5675--5715 & $<0.87$        & \ion{O}{iii}    & 5550--5620 & $2.72\pm0.54$\\
    \ion{He}{ii}     & 4686 & 4676--4696 & $<0.76$        & \ion{C}{iii}     & 5785--5840 & $<0.99$        & \ion{O}{i}?     & 8180--8260 & $2.90\pm0.47$\\
    \ion{He}{i}      & 4713 & 4703--4723 & $<0.78$        & \ion{C}{ii}?     & 6040--6120 & $<1.13$	   &                 &            & \\   
    \ion{He}{i}      & 4921 & 4911--4931 & $<0.75$        & \ion{C}{ii}      & 6130--6180 & $<0.87$ 	   &                 &            & \\   
    \ion{He}{i}      & 5015 & 5005--5025 & $<0.78$        & \ion{C}{ii}      & 6550--6600 & $<0.94$        &                 &            & \\   
    \ion{He}{ii}     & 5411 & 5401--5421 & $<0.69$        & \ion{C}{iii}     & 6700--6760 & $<0.86$        &                 &            & \\   
    \ion{He}{i}      & 5876 & 5866--5886 & $0.80\pm0.21^\mathrm{a}$ & \ion{C}{ii}      & 6760--6820 & $1.79\pm0.29$  &                 &            & \\   
    \ion{He}{i}      & 6678 & 6668--6688 & $<0.49$        & \ion{C}{ii}      & 7210--7260 & $4.67\pm0.32$  &                 &            & \\   
    \ion{He}{i}      & 7065 & 7055--7075 & $<0.50$        & \ion{C}{iii}?    & 7700--7750 & $1.16\pm0.36$  &                 &            & \\
    \ion{He}{i}      & 7286 & 7276--7296 & $<0.67$        &                  &            &                & \multicolumn{3}{l}{$^\mathrm{a}$ Possibly affected by telluric absorption} \\
    \hline
  \end{tabular}
\end{table*}

In Table~\ref{tab:ew}, we list detections and upper limits on the
equivalent widths of several hydrogen and helium lines, which are
routinely seen in optical spectra of classical X-ray binaries and
cataclysmic variables. We do not detect any significant (upto
$3\sigma$) hydrogen and helium emission or absorption lines. The
\ion{He}{i} at 5876\,\AA\ feature is detected at $4\sigma$, though
this measurement may be affected by telluric absorption. Equivalent
widths were also determined for spectral ranges specified in
\citet{njmk04}. These ranges can be attributed to features of carbon
and oxygen, and several of these features are detected.

Before each spectral observation, a $B$ or $R$-band acquisition image
was obtained. These images were bias-substracted and flat-fielded and
instrumental magnitudes for stars on these images were determined
through point-spread-function (PSF) fitting. The instrumental
magnitudes were calibrated against the $BVRI$ observations described
in \citet{bjzv06}. We find that on April 29, 2006, the optical
counterpart to \bron\ had $B=19.89\pm0.03$, which is
$0.22\pm0.04$\,mag fainter than during the observation in March 2000
\citep{bjzv06}. The $R$-band magnitudes are $R=19.52\pm0.05$ on May 5,
2006 and $R=19.55\pm0.03$ on January 1, 2007, which are again
$0.22\pm0.04$\,mag fainter than during the 2000 observation, in
accordance with the higher X-ray flux found in 2000 than in 2007 (see
Fig.~\ref{rxte}).

\section{Discussion}
\label{discussion}

The optical spectroscopic data do not show features from hydrogen. We
regard this as support of the UCXB proposition by \citet{bjzv06}. Time
histories of X-ray or optical flux failed to find periodicities that
would unambiguously confirm the ultracompact nature through a
measurement of the orbital period.

The spectral analysis of the 4 good-quality X-ray spectra finds that
the 0.1-100 keV unabsorbed flux varies between 2.4 and
8.9$\times10^{-10}$~\ecs.  The ASM data provide the best measurement
of the {\it average} persistent flux. If we scale the spectral results
of the low-flux XMM-Newton spectrum to this flux and assume the same
comptonized spectrum, \bron\ on average shines at
$2.6\times10^{-10}$~\ecs\ (0.1-100 keV; unabsorbed). If the peak
bolometric flux of the X-ray bursts ($\approx10^{-7}$~\ecs, see
Fig.~\ref{wfcbursts}) is equal to the Eddington limit \citep[this
statement is accurate to perhaps 10\%; e.g.,][]{gal07}, then the
average persistent flux translates to 0.26\% of the Eddington limit
and the maximum flux to 0.9\%.  These numbers have a limited accuracy
due to the extrapolation to 0.1-100 keV from the 0.3 to 20 keV
bandpass (accurate to roughly a factor of 2; see power-law and
comptonized-spectrum results for the XMM-Newton spectra in
Table~\ref{tabfit2}) and the uncertainty in the inclination angle of
the accretion disk. For an inclination angle of 90\degr\ the observer
views the disk from the side and will measure a low, if not
negligible, accretion flux. There are no signs of dips nor eclipses in
the X-ray flux, implying that the inclination angle $i$ is at most
70\degr\ \citep{horne85}. Thus, the correction factor to derive the
accretion flux from the observed flux is at most 3
(=1/cos70\degr). Combining both uncertainties, the ratio between the
persistent flux and the Eddington limit is accurate to within a factor
of 4. This is a constraining measurement since it is significantly
below the 2\% threshold, a regime in which persistent sources are
expected to be UCXBs. Therefore, this measurement is consistent with
the UCXB nature of \bron. The 2\% threshold is empirically determined
\citep[of all LMXBs with measured orbital periods, only UCXBs occupy
this regime][]{zan07} and qualitatively explained by the smaller
accretion disks in UCXB needing a smaller irradiation flux to remain
in a hot high-viscosity state that sustains accretion onto the compact
object
\citep{jvp96}.

The spectral changes of \bron\ (see Fig.~\ref{pcaspectra}) are similar
to what is observed in other LMXBs \citep[e.g.,][]{gla07,don07}. One
explanation is that at the lowest mass accretion rates the inner disk
is truncated, leaving a hot inner flow with a hot optical thin
boundary layer \citep[][]{mey94,nar94}. As the accretion rate
increases, the disk moves in and changes the hot (high k$T_{\rm e}$)
optically thin (low $\tau$), geometrically thick inner flow to a
cooler (low k$T_{\rm e}$), optically thick (high $\tau$),
geometrically thin disk \citep[e.g.,][]{esin97}. The $L_{\rm
bol}/L_{\rm Edd}$ ratio at which this happens ($\sim1$\%) is roughly
consistent with observations of other LMXBs that descends that far in
luminosity \citep{gla07}. If the accretion rate would have approached
the Eddington limit, presumably $\tau$ would have become so large that
the comptonized spectrum transforms into a black body spectrum. That
does not happen in \bron. One should note that our bandpass is
limited. There may be additional spectral components above our 20 keV
bandpass limit due to comptonization elsewhere in the binary
\citep[e.g.,][]{don07,pai06} or below 0.3~keV in the form of a cool
disk black body. Finally, we remark that the spectra do not seem to
suggest that UCXBs behave differently from non-ultracompact LMXBs. The
fact that UCXBs generally have harder spectra \citep[e.g.,][]{zan07}
is likely because they reside at low enough accretion rates for a
longer fraction of the time. Non-ultracompact LMXBs are only in that
regime when they are transient and in the final decay phase of an
outburst.

We find a larger-than-solar Ne/O abundances ratio. The ratio is 62\%
higher than solar with a significance of 4$\sigma$. High Ne/O ratios
have also been measured from other UCXBs employing XMM-Newton and
Chandra \citep{jc03}. However, it was found that the Ne/O ratio tends
to change from observation to observation with accompanying changing
continuum fluxes. Sometimes those changes would make the value
consistent with the solar value. This was interpreted \citep{jc05} as
the result of different ionization effects on Ne and O, which would
nullify the use of the Ne/O ratio as a possible diagnostic for the
composition of the donor atmosphere. The same appears to be the case
in \bron: the {\it Swift-}XRT spectrum shows a different Ne abundance,
pointing to a different Ne/O ratio.

From a comparison of 0.1-100 keV BeppoSAX spectra of 10 LMXBs in
globular clusters, \citet{sid01} found a spectral diagnostic to
discriminate ultracompact cases against non-ultracompact ones. In a
model consisting of a disk black body and a comptonized component,
only in UCXBs does the inner temperature of the disk black body
resemble the temperature of the seed photons of the comptonized
component, and only in UCXBs are the inferred disk radii physically
realistic (i.e., equal to what is expected for the radius of the inner
disk edge). We are unable to find evidence for a disk black body
component in \bron, preventing us from verifying this diagnostic in
\bron.

The source \bron\ sofar exhibited 7 type-I X-ray bursts. Most bursts
are intermediately long (i.e., tens of minutes) as would be expected
for UCXB if the accretion rate is less than about 2\% of Eddington
\citep{zan05a,cum06,zan07}. The two latter WFC bursts
(Fig.~\ref{wfcbursts}) are relatively short (a few minutes). These
happen to have occurred during a period when the accretion is showing
enhanced flaring behavior (Fig.~\ref{rxte}). A similar effect is seen
more clearly in 4U 0614+091 (Kuulkers et al., in prep.). The most
likely explanation is that the enhanced flaring is effective in
heating up the neutron star crust to such a level that flashes are
ignited more quickly (within days instead of weeks), flash layers are
thinner (by an order of magnitude), and as a result flash durations
shorter (by the same amount). A quantitative analysis of this effect
may be instrumental in determining worthwhile constraints on the
crust, but the analysis requires larger numbers of bursts to enhance
the statistical significance.

Quick follow-up of one burst with {\it Swift-}XRT allowed us to track
the burst to relatively deep levels and show that the bursts last up
to 30 min with temperatures going down substantially below 1 keV. To
our knowledge, this is the first time that a burst has been tracked
down to these cool regimes so accurately and we find no unexpected
behavior.

The X-ray light curve is typical for an UCXB \citep[c.f.,][]{zan07}: a
tranquil component that changes on a timescale of a year, and a
flaring component with a timescale of a week. This typical behavior
may be related to the probably low $q=M_{\rm donor}/M_{\rm accretor}$
value \citep[see][]{zan07}. A low $q$ value makes the accretion disk
prone to warping and precession \citep[giving rise to for instance the
superhump phenomenon in cataclysmic variables; ][]{whi87,pri96}.
There is another interesting aspect of the PCA light curve
(Fig.~\ref{rxte} bottom panel): the peak levels of the flares decrease
at a pace ($\approx60$\% over 1.5 yr) that is similar to the gradual
decrease of the tranquil level. This suggests that the aspect angle of
the accretion disk to the observer is gradually changing. In other
words, a tilted accretion disk is undergoing a nodal precession with a
period that is larger than the 1.5~yr time span of the observations.
Nodal precession has been confirmed in a number of other X-ray
binaries; famous cases are Her X-1
\citep{katz73} and SS 433 \citep[e.g.,][]{katz80}. Such precession is
thought to be responsible for some super-orbital periods in LMXB light
curves \citep{pried87}. The ASM light curve of \bron\ shows 2 periods
of increased fluxes peaking at roughly 2000 and 2005. Perhaps 5 yr is
a superorbital period. Such a period would be very long in comparison
to other LMXBs \citep[c.f.][]{pried87,wen06} and would imply an
unlikely low donor mass \citep[i.e., less than
$10^{-3}$~M$_\odot$;][]{pat01}, but would not be unprecedented
\citep[GX 3+1 appears to have a 6-yr period, see ][]{hartog03}.  We
are unable to determine the period and verify the nodal-precession
idea any further. An interesting test against it would be a
measurement of the recurrence time of X-ray bursts. For instance, 4U
1820-30 has a half-year super-orbital flux modulation
\citep{pried84}.  This can be attributed to a modulation of the
accretion rate because the X-ray bursts recurrence shows a similar
modulation \citep{chou01,cor03}. The rate of X-ray bursts in \bron\ is
too small to test whether the same behavior applies as in 4U 1820-30.

\acknowledgement We thank Jelle Kaastra and Cor de Vries for useful
discussions, Frank Marshall at NASA-GSFC for investigating the {\it
Swift-}UVOT data on \bron, and an anonymous referee for useful
suggestions that improved the manuscript.  This work is based on
observations obtained with XMM-Newton (an ESA science mission with
instruments and contributions directly funded by ESA Member States and
NASA), ESO telescopes at the Paranal Observatories, {\it Swift}/BAT
transient monitor results provided by the {\it Swift}/BAT team,
RXTE/ASM results provided by the ASM teams at MIT and at the RXTE SOF
and GOF at NASA/GSFC, and BeppoSAX (a joint Italian and Dutch program
that was operational between 1996 and 2002).

  \bibliographystyle{aa}
\bibliography{9361}

\end{document}